\documentstyle[amssymb,preprint,aps]{revtex}
\tightenlines
\begin{document}
\draft
\title{Effect of Director Distortions on Polymer-Liquid Crystal Phase
Separation}
\author{D. Voloschenko, O. P. Pishnyak, S. V. Shiyanovskii, and O. D.
Lavrentovich}
\address{Chemical Physics Interdisciplinary Program and Liquid Crystal\\
Institute, Kent State University, Kent, OH 44242, USA}
\date{\today }
\maketitle

\begin{abstract}
We study photopolymerization in a low-molecular weight liquid crystal with
non-uniform director gradients. Phase separation results in spatially
non-uniform distribution of polymer density controlled by the distorted
nematic matrix. Director-gradient-controlled polymerization provides a new
and useful technique to assemble micron-scale polymer architectures.
\end{abstract}


Composites of polymers and liquid crystals (LCs) are subject of intensive
studies for electrooptical applications \cite{CrawfordZumer-book}. The most
popular method to produce composites such as polymer-dispersed LCs (PDLCs)
and polymer-stabilized LCs (PSLCs) is ptotoinduced phase separation in a
mixture of a non-reactive LC and a reactive monomer. Morphology of the
product depends on many factors, such as solubility \cite{Dierking-APL97},
diffusivity \cite{Hikmet-Nature98}, concentration \cite{Amundson} and
reactivity of components, spatial distribution of light intensity \cite%
{Bunning-ARMS2000}, electric field gradients \cite{West-APL98}, time scales
of polymerization \cite{Lapena-PRE99,Vorflusev,Nephew}, etc. Phase
separation from the LC continuous state is of especial interest, as the
morphology can be influenced by the orientational order of the LC solvent. \
For example \cite{loudet}, phase-separating droplets of isotropic fluid form
chains parallel to the director ${{\bf \hat{n}}}$, which specifies the
average orientation of LC molecules. \ In PSLCs, polymer strands can align
along ${{\bf \hat{n},}}$ especially when formed from a mesogenic monomer %
\cite{CrawfordZumer-book,Held}. However, it is not clear how the morphology
would be altered if $\ {{\bf \hat{n}}}$ is distorted and these distortions
are non-uniform in space. Experiments on periodically-distorted cholesteric
gratings demonstrate that the phase-separating component has density
modulation with the period set by the modulated ${{\bf \hat{n}}}$ \cite%
{Vol-ODL-APL99,Kang-APL2000}. There are at least three possible mechanisms
of this effect. \ (1) Since the LC is an elastic medium, the phase
separating component might be forced to accumulate in the sites with the
highest energy of director distortions, especially when this component forms
compact particles rather than fibrils. (2) Polymer density might vary
because of spatial changes of the angle between light polarization ${\bf P}$
and ${{\bf \hat{n}}}$. (3) Variations in light intensity caused by `lens'
effects in the distorted birefringent LC.

In this work, we separate the three mechanisms and demonstrate that spatial
variation of the director gradients can indeed lock-in the spatial
distribution of the precipitating component.

We use the nematic E7 (EM Industries) with a small ($\approx $5wt\%) amount
of the UV-curable prepolymer NOA65 comprised of trimethylol-propane diallyl
ether, trimethylol-propane tris(thiol), isophorone diisocyanate ester,
premixed with a benzophenone photoinitiator \cite{NOA65}.
Photopolymerization proceeds by cross-linking reactions and results in
compact morphologies with a reduced interfacial area between NOA65 and LC
domains; LC is expelled from the polymerized NOA65 \cite%
{West-APL98,Vorflusev,Nephew,Smith-PRL93,Lovinger,Bhargava,Nwabunma-Macromol98}%
. The latter feature makes NOA65 attractive for PDLCs in which LC droplets
are well separated by the polymer matrix. Because of a small concentration
of NOA65, our system is an ''reverse'' of PDLCs. The non-cured mixture
preserves the nematic state at room temperature and upon heating to 53 C$^{0}
$ \cite{Smith-PRL93}. We used a narrow-band UV lamp (wavelength $\lambda $ =
366 nm) and a Glan-Thompson prism to polarize UV light linearly; light
intensity after the polarizer was $I\approx 30$ $\mu $W/cm$^{2}$.
Experiments were performed at room temperature for the cells of thickness $d=
$20 $\mu $m. \ First, we show that the polymerization rate depends on the
angle between ${\bf P}$ and ${{\bf \hat{n}}}$.

{\bf Light polarization mechanism.} The plates of LC cells were treated for
planar alignment (rubbed polyimide layers), ${{\bf \hat{n}=}}$const. \
Photopolymerization was induced by UV light with ${\bf P\bot }{{\bf \hat{n}}}
$ \ or ${\bf P\Vert }{{\bf \hat{n}}}$. Comparison of otherwise identical
cells shows that polymerization is faster when ${\bf P}\Vert {{\bf \hat{n}}}$
(Fig.1). The first few hours of irradiation do not cure NOA65 completely: \
if irradiation is stopped and the sample is heated to 35-50 C$^{0}$, then
the phase-separated NOA65-rich particles dissolve back into the nematic
matrix. \ Note also that the NOA65 particles often form chains along the
groves of rubbed polyimide; as discussed in Ref.\cite{Muthukumar}, patterned
substrates can influence the phase-separating morphologies.

{\bf Director inhomogeneity.} To elucidate the role of director gradients on
morphology of phase-separation, we studied 90$^{\circ }$ twisted cells. The
rubbing directions at the two plates of the cell are perpendicular to each
other. The equally probable domains of opposite 90$^{\circ }$ twists, $%
(n_{x},n_{y},n_{z})=(\cos \pi z/2d,\sin \pi z/2d,0)$ and $%
(n_{x},n_{y},n_{z})=(\cos \pi z/2d,-\sin \pi z/2d,0)$ (the $z$ axis is
perpendicular to the cell plates), are separated by defects (disclinations %
\cite{Geurst}).

The cell configuration allows us to separate the lens and light polarization
effects from the effect of director distortions, by chosing ${{\bf P||\hat{n}%
}}$ at the plate of incidence, Fig.2,3. Since $d>>\lambda $, ${\bf P}$
rotates with ${{\bf \hat{n}}}$ remaining parallel to ${{\bf \hat{n}}}$
(Mauguin regime) within the twist domains. Therefore, the polarization
effect facilitates polymerization within the twist domains but not at the
domain boundary defect, where ${\bf P}$ is not parallel to ${{\bf \hat{n}}}$%
, because the Mauguin regime is violated \cite{Geurst}. Furthermore, light
with ${\bf P}\Vert {{\bf \hat{n}}}\ $propagates with a refractive index $%
n_{e}$, which is the maximum value of the refractive index in the optically
positive E7; thus if the domain boundary acts as a lens, this lens can only
be divergent. Therefore, both polarization and lens effects facilitate NOA65
precipitation within the domains but not at the domain boundaries. \

In the experiment, the NOA65-rich particles start to appear everywhere,
within the twist domains and at the domain boundaries. \ However, many
particles that appear within the twist domains near the boundaries migrate
toward the boundaries and are pinned there, Figs. 2,3. \ Smaller particles
might coalesce into chains of big particles or even form continuous walls.

Qualitatively, the leading mechanism of aggregation in the regions with the
highest LC distortions can be explained as follows. Initially, the nematic
order is preserved in the whole specimen and the NOA65 monomers are
distributed homogeneously \cite{SVS-PRE94}. A phase-separated NOA65 particle
of a radius $R_{p}$ eliminates the nematic order in the volume ~${R_{P}^{3}}$%
. If ${{\bf \hat{n}}}$ were uniform, there would be no preferable site for
the particle (we do not consider here the surface-mediated heterogeneous
nucleation at the cell plates). However, in a distorted LC, elimination of
the volume ${R_{P}^{3}}$ decreases the elastic energy by $E_{V}={{%
KA^{2}R_{P}^{3}/\xi ^{2}}}$, where $K$ is the average elastic constant, $A$
is the amplitude and $\xi $ is the characteristic length of director
distortions. The higher the distortions ${A/\xi }$, the higher the energy
gain $E_{V}$ of replacing the LC region with a particle.

There are other factors, such as director anchoring at the particle surface,
which would act differently depending on whether $R_{p}$ is smaller or
larger than the length $L_{p}=K/W_{p}$, usually of the order of $\left(
0.1-10\right) $ $\mu $m; $W_{p}$ is the anchoring coefficient at the
particle-nematic interface. When $R_{p}>$ $L_{p}$, ${{\bf \hat{n}}}$ around
the particle is determined mainly by the particle's surface propeties; the
gain $E_{V}$ might be larger or smaller than the energy $E_{ind}$ of
additional distortions induced around the particle. There are cases when $%
E_{V}>E_{ind}$ undoubtfully. For example, a spherical particle with
perpendicular anchoring of ${{\bf \hat{n}}}$ \ placed in the LC with a
radial director (point defect-hedgehog) would find the energy minimum at the
center of the defect rather than in any other location, as clearly
demonstrated for nematic colloids with water droplets \cite{Poulin-PRE98}. \
It is, however, hard to find a universal description of $E_{ind}$ and $E_{V}$
when $R_{p}>$ $L_{p}$. Fortunately, for $R_{p}<$ $L_{p}$ (initial stages of
polymerization), it is possible to show that the anchoring effects are not
essential, and that $E_{V}$ dominates to guide the particles toward the
cites with the highest director gradients. \

We use the standard free energy functional

\begin{equation}
F=\int\limits_{V}{f\left( {x,z}\right) }dV+%
%TCIMACRO{\tfrac{1}{2}}%
%BeginExpansion
{\textstyle{1 \over 2}}%
%EndExpansion
\int\limits_{S}{W\left( {1-\left( {{\bf \hat{\nu}}\cdot {\bf \hat{n}}}%
\right) ^{2}}\right) }dS,
\end{equation}
\[
2f={K_{1}}\left( {div{\bf \hat{n}}}\right) ^{2}+{K_{2}}\left( {{\bf \hat{n}}%
\cdot curl{\bf \hat{n}}}\right) ^{2}+{K_{3}}\left[ {{\bf \hat{n}}\times curl%
{\bf \hat{n}}}\right] ^{2},
\]
where $K_{1}$, $K_{2}$, and $K_{3}$ are the elastic constants of splay,
twist, and bend, respectively; ${{\bf \hat{\nu}}}$ is the normal to the
polymer-liquid crystal interface. The change in the free energy caused by
placing a polymer particle in a distorted LC is
\begin{equation}
\Delta E=-E_{V}+\Delta E_{S}+E_{ind},  \label{dE}
\end{equation}
where $E_{V}=\int\limits_{V_{P}}{f\left( {x,z}\right) dV_{P}}$ is the
elastic energy of the LC volume ${V_{P}}${\ that is} excluded by the
particle; $\Delta E_{S}$ is the difference in the anchoring energy at the
surface $S_{P}$ of the particle for homogeneous and distorted director
fields; $E_{ind}$ is the elastic energy of the additional distortions
induced by the particle. For $R_{p}<L_{p}$,

\begin{equation}
E_{V}\cong \frac{1}{2}K\int\limits_{V_{P}}{\left( {\nabla {\bf \hat{n}}}%
\right) ^{2}}dV_{P}\cong \frac{2}{3}\pi K\frac{{R_{p}^{3}}}{{\xi ^{2}}}A^{2}
\label{Ev}
\end{equation}
dominates in Eq. (\ref{dE}), because $\ \Delta E_{S}=$ $\frac{{4\pi }}{{15}}%
W_{p}R_{P}^{4}{div}\left( {{\bf \hat{n}}_{0}div{\bf \hat{n}}_{0}+{\bf \hat{n}%
}_{0}\times curl{\bf \hat{n}}_{0}}\right) $ $\cong \frac{{4\pi
W_{p}R_{P}^{4}A^{2}}}{{15\xi ^{2}}}$ $\cong \frac{{2R_{P}}}{{5L_{P}}}E_{V}$ $%
<<E_{V}$ \ (${\bf \hat{n}}_{0}$ is the director at the center of the LC
region replaced by the particle) and, according to \cite{Kuksenok-PRE96}, $%
\left| E_{ind}\right| \cong \frac{{W_{p}R_{P}^{{}}}}{{5A^{2}K}}\left| \Delta
E_{S}\right| \leqslant \left| \Delta E_{S}\right| $. The tendency of
particles to occupy the regions with high elastic free energy prevails over
the entropy driven randomization if $\left| \Delta E\right| \approx
E_{V}\geqslant k_{B}T$ and, therefore, $R_{P}\geqslant R_{C}=\sqrt[3]{{\frac{%
{k_{B}T\xi ^{2}}}{{2KA}^{2}}}}$. For $K=10^{-11}N$, $\xi =5\mu m$, $A=1$,
and room temperature, one finds $R_{C}=0.2\mu m$. To estimate time $\tau $,
required to drive the particle, one can use the Stokes formula for the drag
force $F\cong {\Delta E/}\xi \cong 6\pi \alpha _{4}R_{P}^{{}}\bar{v}$, where
$\alpha _{4}$ is the Leslie viscosity coefficient and $\bar{v}\cong \xi
/\tau $ is the average velocity of the particle. Substitution $\left| \Delta
E\right| \approx E_{V}$ in the Stokes formula results in $\tau \approx \frac{%
{9\alpha _{4}\xi ^{4}}}{{KA^{2}R_{P}^{2}}}$. For $\alpha _{4}=0.1$ Pa$\cdot $%
s, $\xi =5\mu m$, $R_{P}=1\mu m$, and $A=1$, the driving time $\tau \approx
50$ $s$ \ is small enough to provide accumulation of NOA65-rich particles.
However, its strong dependence on $\xi $, $\tau \propto \xi ^{4}$, shows
that very smooth distortions (large $\xi $) may not be sufficient for
spatial inhomogeneity of phase-separating NOA65.

{\bf Regular morphologies.} Regularly distorted director patterns can be
used to fabricate regular polymer architectures, as we show below for the
``fingerprint'' cholesteric textures in cells with homeotropic alignment
(lecithin coatings). The cholesteric LC (pitch $p=10\mu $m) was obtained by
adding a chiral agent CB15 (EM Industries) to E7/NOA65 mixture. The
E7/CB15/NOA65 mixture was doped with a fluorescent dye acridine orange base
(0.1 wt \% ) for confocal polarizing microscope (Olympus Fluoview BX-50)
observations. Confocal microscopy allows one to obtain thin (micron scale)
optical slices of the sample both in the horizontal and vertical planes,
see, e.g., Refs. \cite{Amundson,Nephew,Held,Smalyukh-CPL}.

Before phase separation, the dye is distributed homogeneously \cite%
{SVS-PRE94}, and the confocal polarizing microscope images the director
field, since the anisometric dye molecules are oriented by ${{\bf \hat{n}}}$%
. Different orientations result in different intensity of fluorescence \cite%
{Smalyukh-CPL}: minimum when ${{\bf \hat{n}}}$ is along the $z$-axis and
maximum when ${{\bf \hat{n}}}$ is parallel to the polarization of the
probing light. The vertical $(xz)$ cross-section has a characteristic
diamond-like shape, with the axis slightly tilted away from the $z$-axis,
Fig. 4a. The confocal image corresponds to the computer-simulated
equilibrium director ${{\bf \hat{n}}}(x,z)$, Fig. 4b. With the known ${{\bf
\hat{n}}}(x,z)$, one can map the local energy density $f(x,z)$, Eq. (1) with
the added twist term $2\pi {K_{2}}\left( {{\bf \hat{n}}\cdot curl{\bf \hat{n}%
}}\right) /p$. The energy pattern, Fig. 4c, shows that subsurface regions
are distorted stronger than the bulk of the cell. It is easy to understand,
since in the bulk, ${{\bf \hat{n}}}$ has a favorable helical structure,
whereas near the boundaries, energetically costly bend and splay occur to
satisfy the perpendicular boundary condition.

The morphology of phase-separation induced by nonpolarized UV-light is
modulated in accordance with the fingerprint structure. In the phase
separated system, the acridine dye tends to accumulate in the polymer
component rather than in the LC \cite{Tahara-SPIE98}. In the confocal
textures, Figs. 4d,e, the stained NOA65-rich particles appear bright on the
dark LC background. Comparison of thin horizontal $(x,y)$ optical slices
near the bounding plates reveals that the particles are located exactly in
the places with the maximum energy of director distortions, Fig. 4c-e (these
confocal microscopy data are also confirmed by our scanning-electron
microscopy observations). Light polarization mechanism also helps to
accumulate NOA65 in these places, even when light is not polarized, since $%
n_{z}<1.$

{\bf Summary. \ }The director field can influence the pattern of phase
separation from the LC solvent: the regions with the strongest director
distortions can attract the phase-separating component. \ The effect
demonstrates a principal difference between phase separation in isotropic
and ordered anisotropic solvents. \ Similar effect might be expected even
when the initial state of an anisotropic solvent is uniform:
phase-separating particles that reach a characteristic size $L_{p}$ will
distort the director field and thus influence the phase separation around
them. Further studies are needed to elucidate how the morphologies will be
altered by such factors as the nature of monomer (e.g., mesogenic vs.
non-mesogenic), diffusivity, reactivity, etc. We also demonstrated that the
rate of NOA65 photopolymerization in a nematic LC is higher when the
polarization of UV light is parallel to the director.

We thank L.-C. Chien, P. Palffy-Muhoray, S. Sprunt, J.L. West and D.-K. Yang
for usefull discussions. \ The work has been supported by the NSF ALCOM
grant, DMR 89-20147 and by donors of the Petroleum Research\ Fund,
administered by ACS, grant 35306-AC7.

\bigskip

{\bf Figure Captions\bigskip }

FIG. 1. Polymerization dynamics in E7/NOA65. \ Microphotographs are taken
after 35 min of UV exposure with (a) ${\bf P\bot }{{\bf \hat{n}}}$ and (b) $%
{\bf P}\Vert {{\bf \hat{n}}}$. Bright spots represent NOA65-rich particles.
(c) $S/S_{0}$ vs irradiation time for ${\bf P\bot }{{\bf \hat{n}}}$
(triangles) and ${\bf P}\Vert {{\bf \hat{n}}}$ (circles); $S$ is the area
occupied by precipitated particles, as seen under the microscope, $S_{0}$ is
the whole area of CCD frame.

FIG. 2. UV photopolymerization near the (horizontal) defect boundary
separating two twist domains in E7/NOA65. Two NOA65 particles marked by a
circle and a square migrate to the boundary. \ Double arrow shows the
directions of both ${\bf P}$ and ${{\bf \hat{n}}}$ at the cell plate facing
the UV\ lamp. \ Time bars show the UV exposure time. Polarizing microscopy
with uncrossed polarizers.

FIG. 3. Phase-separated E7/NOA65 near a defect boundary between two twist
domains. Exposure time 1 hr. Uncrossed polarizers.

FIG.4. Phase separation in a homeotropic cell ($d=10$ $\mu m$) with
cholesteric ''fingerprints''. (a), (b), (c) are the vertical cross sections
of the cell, and (d) and (e) are the horizontal one. \ (a) confocal texture
of ${{\bf \hat{n}}}\left( x,z\right) $; (b) computer simulated ${{\bf \hat{n}%
}}\left( x,z\right) $; $K_{1}$ = 6.4, $K_{2}$= 3, $K_{3}$= 10 (all in pN), $%
W_{c}d/K$= 10; $W_{c}$ is the anchoring coefficient at the plates; (c) map
of $f\left( x,z\right) $, higher energy densities are darker. Confocal
textures of polymer particles near the upper plate (d) and the lower plate
(e).

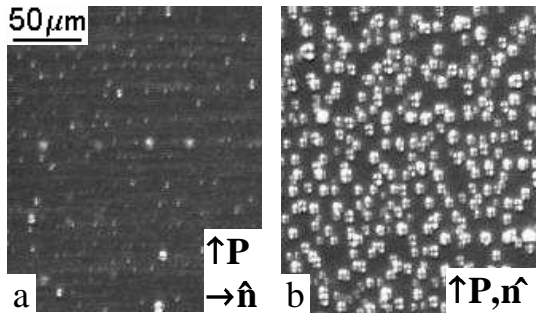
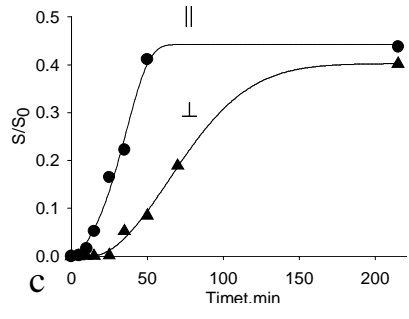
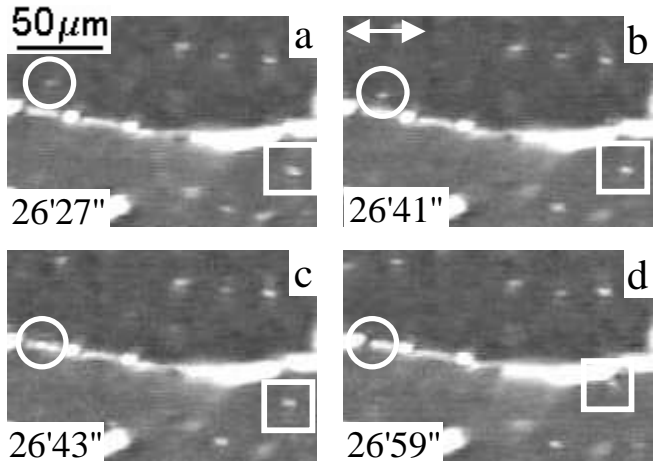
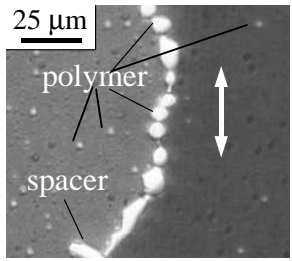
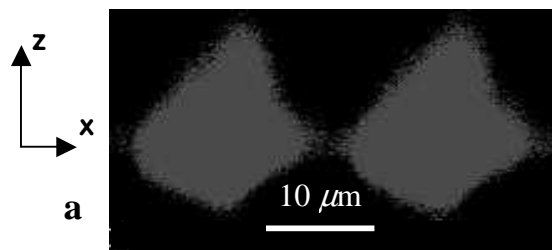
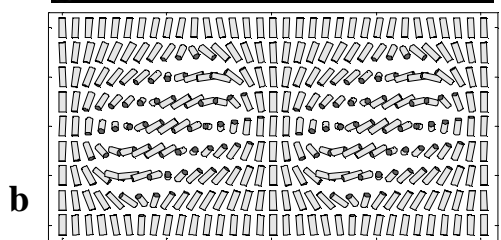
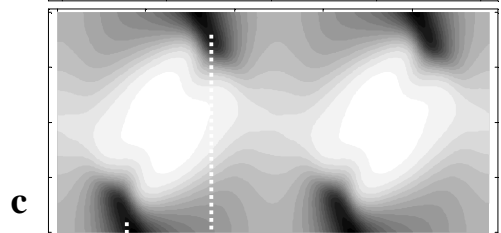
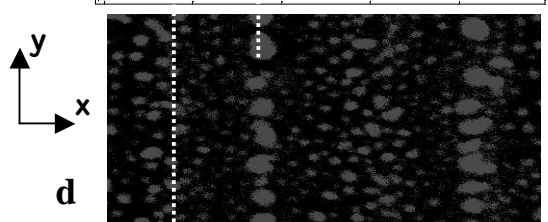
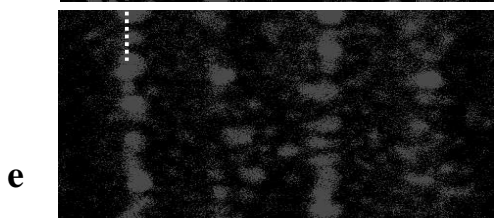
\end{document}